\documentclass[dvips,ejs,preprint]{imsart}

\RequirePackage[OT1]{fontenc}
\RequirePackage{amssymb,amsthm,amsmath}
\RequirePackage{natbib}
\RequirePackage[colorlinks,citecolor=blue,urlcolor=blue]{hyperref}

\usepackage{graphicx}
\usepackage{multirow}
\arxiv{arXiv:1203.3422}

\startlocaldefs
\numberwithin{equation}{section}
\theoremstyle{plain}
\newtheorem{theorem}{Theorem}[section]
\newtheorem{proposition}{Proposition}[section]
\def\RR{\mathbb{R}}

\def\NN{\mathbb{N}}

\def\EE{\mathbb{E}}
\def\II{\mathbb{I}}
\def\dd{\mathrm{d}}
\def\ee{\mathrm{e}}
\endlocaldefs

\begin{document}

\begin{frontmatter}
\title{Statistics for the Luria-Delbr\"uck distribution}
\runtitle{Statistics for the Luria-Delbr\"uck distribution}

\begin{aug}
\author{\fnms{Agn\`es} \snm{Hamon}
\ead[label=e1]{Agnes.Hamon@imag.fr}}
\and
\author{\fnms{Bernard} \snm{Ycart}
\ead[label=e2]{Bernard.Ycart@imag.fr}}

\runauthor{A. Hamon and B. Ycart}

\affiliation{Universit\'e de Grenoble and CNRS}

\address{Laboratoire Jean Kuntzmann\\
Universit\'e de Grenoble and CNRS\\ 
51 rue des Math\'ematiques 38041 Grenoble cedex 9, France}

\end{aug}

\begin{abstract}
The Luria-Delbr\"uck distribution is a classical model of mutations
in cell kinetics. It is obtained as a limit when 
the probability of mutation tends to zero and the number 
of divisions to infinity. It can be interpreted as a compound Poisson
distribution (for the number of mutations) of exponential mixtures
(for the developing time of mutant clones) of geometric distributions
(for the number of cells produced by a mutant clone in a given time).
The probabilistic interpretation, and a rigourous proof of
convergence in the general case, are deduced from classical results on
Bellman-Harris branching processes. The two parameters of the
Luria-Delbr\"uck distribution are the expected number of mutations,
which is the parameter of interest, and the relative fitness of normal
cells compared to mutants, which is the heavy tail exponent. 
Both can be simultaneously estimated by the maximum
likehood method. However, the computation becomes numerically unstable 
when the maximal value of the sample is large, which
occurs frequently due to the heavy tail property. Based on the
empirical probability generating function, 
robust estimators are proposed 
and their asymptotic variance is given. 
They are comparable in precision to maximum likelihood
estimators, with a much broader range of calculability, a better
numerical stability, and a negligible computing time. 
\end{abstract}



\end{frontmatter}

\section{Introduction}
\cite{LuriaDelbruck43} reported an experiment on virus resistant 
bacteria: cultures of the same strain of \emph{Escherichia Coli} 
having grown up independently for several generations,
the cells were plated onto selective medium and surviving bacteria counted.
The major feature of the data was a surprisingly high fluctuation, 
with frequent appearance of large counts.
This experiment, adapted since on many different types of cell cultures, 
founded \emph{fluctuation analysis}, whose objective is to estimate 
probabilities of mutations. Mathematical models were introduced by 
\cite{LeaCoulson49}, and Bartlett in the discussion following 
\cite{Armitage52}: see chap.~II p.~59 of \cite{Kendall52}  
for an early review, and \cite{Zheng99,Zheng10} for more recent
ones. Classical modeling hypotheses are the following:
\begin{itemize}
\item at time $0$ a homogeneous culture of $n$ normal cells is given;
\item the generation time of any normal cell is a random variable with
  distribution function $G$;
\item when a division of a normal cell occurs, it is replaced by:
\begin{itemize}
\item one normal and one mutant cell with probability $p$,
\item two normal cells with probability $1-p$;
\end{itemize}
\item the generation time of any mutant cell is exponentially
  distributed with parameter $\mu$;
\item when a division of a mutant cell occurs, it is replaced by two
  mutant cells;
\item all random variables and events (division times and mutations) 
are mutually independent.
\end{itemize}
In Kendall's notations \cite[p.~61]{Kendall52}, the model considered
here is the G/M/0 (general distribution for normal cells, Markovian
evolution of mutants, no phenotypic lag).
The particular case where generation times of normal cells follow the
exponential distribution (M/M/0 or ``fully stochastic'' model), appeared for
the first time in the discussion following  
\cite[p.~37]{Armitage52}. There Bartlett obtained the asymptotics for
equal growth rates ($\mu=\nu$) for large $n$ and $t$ and small $p$; he
later generalized to differential 
growth rates \cite[p.~134]{Bartlett78}.  
In between,  \cite{Mandelbrot74} had proposed similar
asymptotics, but the paper had several errors (pointed out by
\cite{Pakes93} and \cite{Zheng02}), and no mathematical
proof nor interpretation was given (see \cite{Zheng08}). Several
extensions of the M/M/0 case have been proposed, notably by 
\cite{OpreaKepler01} and \cite{Angerer10}. Yet,
to the best of our knowledge, no rigourous
proof of convergence, and no clear probabilistic interpretation has
been given for the G/M/0 model. 
\vskip 2mm\noindent
Our first objective is to establish the convergence 
in distribution of the number of mutants 
in the general case, and give a probabilistic interpretation
of the result, based on the theory of branching processes.
To state our main result, we need to introduce the growth rate $\nu$
and the proportionality constant $C$ associated to the distribution
$G$, for a binary division dynamics in a Bellmann-Harris process
(\cite{Harris63,AthreyaNey72}). The growth rate $\nu$ (also called
\emph{Malthusian parameter}) is  defined as the unique root of the equation:
\begin{equation}
\label{malthus}
2\int_0^{+\infty} \ee^{-\nu s}\,\dd G(s)=1\;.
\end{equation}
The proportionality constant $C$ is:
\begin{equation}
\label{propor}
C= \left(4\int_0^{+\infty} s\ee^{-\nu s}\,\dd G(s)\right)^{-1}\;.
\end{equation}
\begin{theorem}
\label{th:main}
Consider the  model G/M/0 described above. Let $p=p_n$ and $t=t_n$ be
two sequences, and $\alpha$ a positive real 
such that:
$$
\lim_{n\to \infty} p_n=0\;,\quad
\lim_{n\to \infty} t_n=+\infty\;,\quad
\lim_{n\to \infty} p_nnC\ee^{\nu t_n} = \alpha\;.
$$
As $n$ tends to $+\infty$, the distribution of the 
number of mutants at time $t_n$, starting with $n$ normal cells at
time $0$, converges to the distribution with
probability generating function:
\begin{equation}
\label{ld}
g_{\alpha,\rho}(z) = \exp(\alpha(h_\rho(z)-1))\;,
\end{equation}
where $h_\rho$ is the probability 
generating function of the Yule distribution with
parameter $\rho=\nu/\mu$:
\begin{equation}
\label{y}
h_\rho(z) = \rho z \int_0^1 \frac{v^\rho}{1-z+z v}\,\dd v\;. 
\end{equation}
\end{theorem}
Following \cite{Zheng02}, we call 
\emph{Luria-Delbr\"uck distribution} with parameters $\alpha$ and
$\rho$, and denote by 
LD$(\alpha,\rho)$, the distribution on the set of integers with
probability generating function $g_{\alpha,\rho}$.
Observe that it depends on $G$ (the generation time distribution of
normal cells) only through its
growth rate $\nu$, thus matching the conclusions of
\cite{Jonesetal93}. 
The parameter $\alpha$ is the mean number of
mutations and $\rho=\nu/\mu$ is the relative \emph{fitness} of normal 
cells compared to mutants. Theorem \ref{th:main} has a very
simple interpretation that 
can be summarized in 3 points (precise justifications from the theory
of branching processes will be given in section \ref{branching}):
\begin{enumerate}
\item
the number of divisions of normal cells before time $t_n$ is equivalent
in probability to
$nC\ee^{\nu t_n}$.  Therefore the expected number of
mutations that can lead to mutants at time $t_n$ is equivalent to
$p_n\times nC\ee^{\nu t_n}\simeq \alpha$.  
Since the number of divisions is large and the probability of mutation
is small, the total number of mutations approximately 
follows the Poisson distribution with parameter $\alpha$, 
by the law of small numbers (this remark had already been made by 
\cite[p.~499]{LuriaDelbruck43});
\item
mutations happen at random instants, but due to exponential growth, the
vast majority of divisions occur rather close to the end of the
observation interval. Actually, the time between the occurrence of a
typical mutation and the end of the observation, asymptotically 
follows the exponential distribution with parameter $\nu$. 
This is the time for which any given mutant
clone (population stemming from a single cell) will develop;
\item
a mutant clone develops according to a Yule process with
parameter $\mu$. Its size at time $s$ is
geometric with parameter $\ee^{-\mu s}$.
\end{enumerate}
Indeed, $h_\rho$ is the probability generating function of an exponential 
mixture of geometric distributions (changing $v$ into $\ee^{-\mu s}$
in (\ref{y})):
\begin{equation}
\label{yp}
h_\rho(z)=\int_0^{+\infty} \frac{z \ee^{-\mu s}}{1-z+z\ee^{-\mu s}}
\,\nu\ee^{-\nu s}\,\dd s\;. 
\end{equation}
Therefore, the Luria-Delbr\"uck distribution is a compound Poisson
of an exponential mixture of geometric distributions. 
It is a heavy tail distribution, with tail exponent $\rho$: the higher the
fitness of mutants compared to normal cells, the heavier the tail. 
This explains the appearance of unusually high counts of mutants in cell
growth experiments. 
\vskip 2mm
The main goal of fluctuation analysis is to estimate the mutation
probability $p$, from a sample of mutant counts. Using theorem
\ref{th:main}, mutant counts can be considered as realizations of the
LD$(\alpha,\rho)$, $\alpha$ and $\rho$ being unknown. If an estimate
of $\alpha$ has been calculated, then an estimate of $p$ can be obtained,
dividing  by the total number of cells at the end of the experiment.
Many methods of estimation for $\alpha$ have been proposed: see
\cite{Foster06}. The simplest consists in estimating the probability
of observing no mutant: $\ee^{-\alpha}$; this is the original
method used by \cite{LuriaDelbruck43}. The relative
fitness is not taken into account, and it can only be used
if $\alpha$ is small. Any other consistent
estimator of $\alpha$ has to be sensitive 
to the value of $\rho$: even if only the parameter
$\alpha$ is of interest, its estimation cannot be decoupled from that
of $\rho$. As far as we know, no thorough statistical
study of the coupled estimation problem has been made. Moreover, even
though fluctuation analysis with differential growth rates has been
advocated by several authors 
(\cite{Koch82}, \cite{Jones94}, \cite{JaegerSarkar95}, and
\cite{Zheng02,Zheng05}), most studies are still being made using  
the LD$(\alpha,1)$ without questionning the equal rate hypothesis
(e.g.  \cite{Wuetal09}, \cite{Jeanetal10}). The main 
objective of this work is to propose statistically founded estimation
procedures for $\alpha$ and $\rho$. 
\vskip 2mm
Maximum Likelihood (ML) seems to be the obvious choice: see
\cite{Zheng05}. Indeed, the likelihood and its derivatives 
can be computed by iterative algorithms: theoretically at least, 
the problem could be considered as solved. This is not so 
in practice, mainly because the multiple sums that must be 
computed by the optimization algorithm make it quite
unstable. According to the numerous tests that we have made, the ML 
estimates cannot be reliably computed for samples whose maximum
exceeds 1000.  
Therefore a robust estimation procedure must be used (see
 \cite{Wilcox12} as a general reference). 
A first approach is Winsorization, that 
consists of replacing any value of the sample that pass a certain
bound, by the bound itself.
As an alternative robust estimation technique, we propose to transform
the data through $X\mapsto z^X$, with $0<z<1$, i.e. consider the
empirical probability generating function. This method, particularly
adapted to compound Poisson distributions, has already been considered
in analogous cases by \cite{RemillardTheodorescu00}
 and  \cite{Marchesellietal08}.
We have derived consistent Generating Function (GF) estimators for
$\alpha$ and $\rho$, for which the asymptotic covariance matrix has
been calculated. They proved to be close to optimal efficiency,
with a broad range of calculability, a good 
numerical stability, and a negligible computing time.
We have developed in R (\cite{R}) a set of functions  that perform the usual
operations on the LD distributions, output ML and GF
estimates, confidence regions and p-values for hypothesis
testing. These functions have been made available 
online\footnote{http://www.ljk.imag.fr/membres/Bernard.Ycart/LD/}.
\vskip 2mm
The rest of the paper is organized as follows. In section \ref{branching}, some
standard results of branching process theory are reviewed, and the
justification of theorem \ref{th:main} is given. 
The implementation of the ML algorithm is described in section \ref{ML}, and its
drawbacks are discussed. Section \ref{GF} is devoted to the 
Generating Function estimators, their asymptotic variance, and their
implementation. The two methods have been compared on a simulation study,
whose results are reported in section \ref{Sim}. 
\section{Bellman-Harris processes}
\label{branching}
This section reviews a few
aspects of supercritical continuous time branching
processes, adapting them to the context of cell division:
see \cite{Harris63,AthreyaNey72} as general references.
The proof of theorem \ref{th:main}, as well as the probabilistic
interpretation of the LD$(\alpha,\rho)$ as a 
compound Poisson of exponential mixtures of geometric distributions,
rely upon the following three assertions (the hypotheses are
those of theorem \ref{th:main}).
\begin{itemize}
\item[A1:]
the number of mutations converges in distribution to 
the Poisson distribution with parameter $\alpha$.
\item[A2:] 
the joint distribution of the developing times of a fixed number $k$
of mutant clones converges in distribution to the product of $k$
independent copies of the exponential distribution with parameter $\nu$.
\item[A3:]
the size at time $s$ of a mutant clone has 
geometric distribution with parameter $\ee^{-\mu s}$.
\end{itemize}
From assertions A2 and A3, the size of any mutant
clone is an exponential mixture of geometric distributions, 
the probability generating function of which is given by (\ref{yp}).
Moreover,
these sizes on a fixed number of mutant clones are asymptotically
independent. From assertion A1, the number of mutants asymptotically
follows a compound
Poisson distribution, hence (\ref{ld}). A first by-product of
this interpretation is a fast simulation algorithm that we have used
extensively for our simulation study.
\vskip 2mm\noindent 
We shall not insist on A3 which is well known:
the geometric distribution of a Yule
process taken at time $s$ is formula (5) p.~35 of 
\cite{Yule25} (see also \cite[p.~109]{AthreyaNey72}). 
Assertion A1 follows from the law of small numbers,
provided we prove that the number of mutation occasions (divisions of
normal cells before $t_n$) is equivalent in probability
to $nC\ee^{\nu t_n}$. Consider
$n$ independent copies of the Bellman-Harris branching process with
generation time distribution $G$ and binary divisions (without
mutations), each starting with one single cell at time $0$. We assume the
usual hypotheses on $G$ to ensure that the number of cells has finite
mean and variance at each instant (see \cite{AthreyaNey72}).  
In each copy, the number of divisions before 
time $t$ is equal to the number of cells living at time $t$ minus
one. Let $N_1(t)$ be the number of cells living 
at time $t$ of one copy. Then:
$$
\lim_{t\to +\infty} \EE[N_1(t)]\ee^{-\nu t} = C\;,
$$
from Theorem 17.1 p.~142 of \cite{Harris63}. 
From there, it is easy to deduce that the 
total number of cells living
at  time $t_n$ for the $n$ independent copies, is equivalent in
probability to $nC\ee^{\nu t_n}$, since $t_n$ tends to infinity.
Hence the total number of divisions in the $n$ copies, $N_n(t_n)-n$,
is also equivalent in probability to the same quantity. 
That number does not have the same distribution as the number of
divisions of normal cells in the G/M/0 model: when a mutation occurs,
the subsequent
mutant clone develops according to a different dynamics. However,
mutations are rare, and the difference between the number of divisions
in the G/M/0 model and in the $n$ independent copies of non mutating
normal clones, remains negligible. To prove that
assertion, a coupling will be constructed. Start with
the $n$ independent copies above, and mark independently each division as
potentially mutant with probability $p_n$, or non mutant with
probability $1-p_n$. Since $p_n nC\ee^{\nu t_n}$
tends to $\alpha$, the number of marked divisions converges in
distribution to the Poisson distribution with parameter $\alpha$,
hence remains bounded in probability. Moreover, with probability
tending to $1$, at most one division per copy has been marked.
To deduce the G/M/0 model from
the marked independent copies, replace clones after divisions marked
as mutants, by the mutant dynamics. Hence with probability tending to
$1$, the number of mutation occasions  in the G/M/0 model is less
than the number of marked divisions in the independent copies, and
differs from it by a bounded number. This proves that the number of
division occasions in the G/M/0 model is equivalent in probability to 
$nC\ee^{\nu t_n}$, as requested.
\vskip 2mm
Assertion A2 can be rephrased into a well known
statement about the empirical distribution of split times in a
continuous time branching process. The split times sequence, 
denoted by $(\tau_i)$, is the increasing sequence of instants 
at which divisions
occur. Actually, a much stronger result than needed here has been
given on the asymptotic 
distribution of the sequence $(\tau_i)$. Theorem \ref{thst} below
\cite[p.~669]{Kuczek82b} states the almost sure convergence of the
empirical distribution of split lags, i.e. the differences $t-\tau_i$,
to the distribution function of the exponential with parameter $\nu$.
\begin{theorem}[\cite{Kuczek82b}]
\label{thst}
As $t$ tends to $+\infty$, and for any fixed $s>0$, the random variable
$$
\left(\sum_{i=1}^{+\infty} \II_{[0,s]}(t-\tau_i)\right)/
\left(\sum_{i=1}^{+\infty} \II_{[0,+\infty)}(t-\tau_i)\right)
$$
converges almost surely to the constant $1-\ee^{-\nu s}$.
\end{theorem} 
The almost sure convergence implies convergence in
distribution: if a split
time $\tau_i$ is chosen at random among those before $t$, then the
distribution of $t-\tau_i$ converges to the exponential with parameter
$\nu$; this is one part of assertion A2. The other part is the
asymptotic independence of a fixed number of split times chosen at
random; it follows from Theorem 3.1 p.~673 of \cite{Kuczek82b}. 
Notice that in the particular case where the generation times
of normal cells are exponentially distributed, assertion A2 is exact,
and not asymptotic.
\begin{proposition}
\label{os}
Let $\{N_t\,,\;t\geqslant 0\}$ be a Yule process and
$(\tau_n)_{n\in\NN}$ its sequence of split times.
Conditionally on $N_t=i+1$, $t-\tau_1,\ldots,t-\tau_i$ are distributed as
$i$ independent random variables ranked in decreasing order, 
each following the exponential distribution with parameter $\nu$,
truncated to $[0,t]$.  
\end{proposition}
This is Reed's statement, from \cite[p.~8]{Reed06}. Different
formulations of the same ``order statistic'' property 
have been given by \cite{Kendall66}
and  \cite{NeutsResnick71}. 
\section{Maximum Likelihood estimators}
\label{ML}
A priori, the Maximum Likekihood method should be the obvious choice
for estimating $\alpha$ and $\rho$: under some mild assumptions it is
asymptotically optimal (\cite{LehmannCasella99}), and  the
asym\-ptotic bias can be improved in different ways (\cite{Eldar06}). 
For the LD$(\alpha,\rho)$, it 
has been recommended by several authors: 
\cite{Maetal92}, \cite{Jonesetal93}, \cite{Zheng02,Zheng05}. Its
main features and drawbacks
are discussed in this section. 
\vskip 2mm
As pointed out by  \cite{Pakes93}, the compound Poisson 
interpretation yields both computation algorithms
and asymptotic results on the LD$(\alpha,\rho)$, even 
though its probabilities do not have an explicit form (see also 
 \cite{Mohle05}, \cite{Dewanjietal05}, and  \cite{Zheng02,Zheng05}). 
Indeed, let $p_k$ denote the probabilities of
the Yule distribution with parameter $\rho=\mu/\nu$ (its probability
generating function being $h_\rho$). For $k\geqslant 1$:
$$
p_k=\int_0^{+\infty} \ee^{-\mu t}(1-\ee^{-\mu t})^{k-1}\,\nu\ee^{-\nu
  t}\,\dd t 
=\rho B(\rho+1,k)\;,
$$ 
where $B$ is the Beta function. 
The probabilities $q_k$ of the LD$(\alpha,\rho)$ can be computed by
the following recursive formula, easily deduced from the probability generating
function (\ref{ld})  
(see \cite{Pakes93} and references therein):
\begin{equation}
\label{algo}
q_0=\ee^{-\alpha}
\;\mbox{and for $k\geqslant 1$, }
q_k = \frac{\alpha}{k} \sum_{i=1}^k ip_i q_{k-i}\;.
\end{equation}
The log-likelihood and its derivatives
with respect to the parameters also have explicit algorithms; they have been
implemented by  \cite{Zheng05}. 

\begin{figure}
\centering
\includegraphics[scale=0.4]{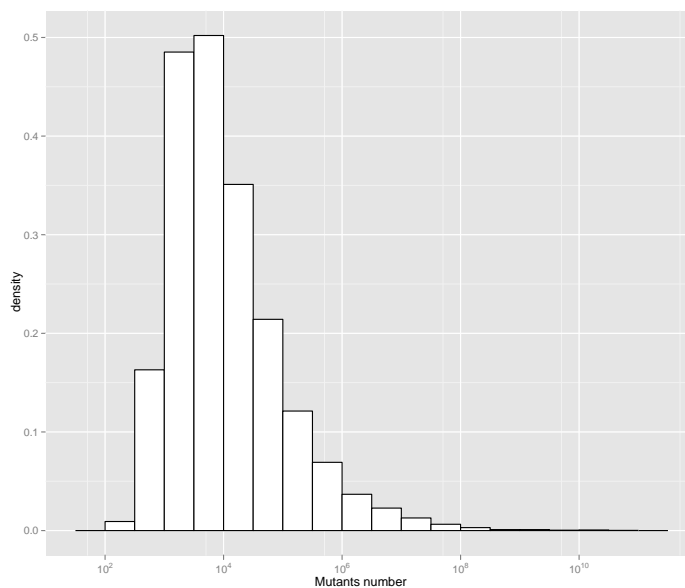}
\caption{Histogram of a $10^5$-sample from the LD$(50,0.5)$ on a
  logarithmic scale. The class intervals are $(10^n\,;\,10^{n+1})$ for
  $n=2$ to $n=15$.}
\label{fig:histogram}
\end{figure}

Let $(X_1,\ldots,X_n)$
be $n$ independent identically  
distributed random variables with common distribution
LD$(\alpha,\rho)$, and $M=\max_j
{X_j}$. For $i\geqslant 0$ we define $c_i=\sum_{j=1}^n 1_{(X_j=i)}$.
The log-likelihood is:
\begin{equation} 
	\ell=\sum_{j=1}^M c_j \log{q_j}.
\end{equation}
 Using (\ref{algo}), the log-likelihood can be calculated iteratively.
Its derivatives are expressed in terms of the derivatives of $q_k$ with
 respect to 
$\alpha$ and $\rho$. These derivatives satisfy the following
equations. For $k\geqslant 1$: 
\begin{equation}
\label{d1}
\begin{array}{lclclcl}
\displaystyle{\frac{\partial q_k}{\partial\alpha}}
&=&\displaystyle{\left(\sum_{h=1}^k
          p_hq_{k-h}\right)-q_k}& \ \text{ and }\ &
\displaystyle{\frac{\partial q_k}{\partial\rho}}&=& 
\displaystyle{\alpha \sum_{h=1}^k
        \frac{\partial p_h}{\partial\rho} q_{k-h}}\;, 
\end{array}
\end{equation}
where the derivative of $p_h$ with respect to $\rho$ can be computed
by a recursive formula (see \cite{Zheng05}) and the derivatives of
$q_0$ in $\alpha$ and $\rho$ 
initialize the algorithms:
\begin{equation}
\label{d0}
\frac{\partial q_0}{\partial\alpha} = -\ee^{-\alpha}\;,\quad
\frac{\partial^2 q_0}{\partial\alpha^2} = \ee^{-\alpha}\;,\quad
\frac{\partial q_0}{\partial\rho} = 
\frac{\partial^2 q_0}{\partial\alpha\partial\rho} =
\frac{\partial^2 q_0}{\partial\rho^2} = 0\;.
\end{equation}
The second derivatives are computed by:
\begin{equation}
\label{d2}
\begin{array}{lcl}
\displaystyle{\frac{\partial^2 q_k}{\partial\alpha^2}}
&=&\displaystyle{\left(\sum_{h=1}^k
  p_h\frac{\partial q_{k-h}}{\partial\alpha}\right)-\frac{\partial
  q_k}{\partial\alpha}}\\[2ex] 
\displaystyle{\frac{\partial^2 q_k}{\partial\alpha\partial\rho}}
&=&\displaystyle{\sum_{h=1}^k
\frac{\partial p_h}{\partial\rho}\left(q_{k-h}+ \alpha \frac{\partial
    q_{k-h}}{\partial\alpha} \right)}\\[2ex] 
\displaystyle{\frac{\partial^2 q_k}{\partial\rho^2}}
&=&\displaystyle{ \alpha\left(\sum_{h=1}^k
  \frac{\partial^2 p_h}{\partial\rho^2} q_{k-h}+\frac{\partial
    p_h}{\partial\rho} \frac{\partial q_{k-h}}{\partial\alpha}
\right)} 
\end{array}
\end{equation}
where the second derivative of $p_h$ with respect to $\rho$ can be expressed
as a function of the Beta, digamma, and trigamma functions. 
The gradient optimization procedure at iteration $i+1$ computes:
\begin{equation}
\begin{pmatrix} \hat{\alpha}_{i+1} \\
  \hat{\rho}_{i+1}\end{pmatrix}=\begin{pmatrix} \hat{\alpha}_i \\
  \hat{\rho}_i\end{pmatrix}-H_i^{-1}D_i\;, 
\end{equation}
where $D_i$ denotes the gradient, and $H_i$ the Hessian of the log-likelihood 
evaluated at $(\hat{\alpha}_i,\hat{\rho}_i)$. So formulas
(\ref{algo}), (\ref{d1}) and (\ref{d2}) must be applied iteratively
for vectors as large as the sample maximum $M$. 
The convolution products involve sums of products of small
terms for large values of $k$, and numerical errors accumulate along
iterations. In practice, the procedure is very long and 
may become numerically unstable as soon as
the maximal value of the sample exceeds 1000. To give an example, the
probability to get at least one 
value larger than 1000 on a 100-size sample of the LD$(2,0.8)$ is $0.53$. 
\vskip 2mm
Moreover, it is quite difficult to calculate dispersion regions, hence
to output confidence intervals and p-values for hypothesis
testing. Indeed, the Fisher information matrix $I(\alpha,\rho)$ is the
following: 
\begin{equation*}
I(\alpha,\rho)=\begin{pmatrix}\sum_{k=0}^{+\infty}
          \left(\frac{\partial q_k}{\partial
              \alpha}\right)^2\frac{1}{q_k}& \sum_{k=0}^{+\infty}
          \frac{\partial q_k}{\partial\alpha}\frac{\partial
            q_k}{\partial\rho}\frac{1}{q_k}\\
\sum_{k=0}^{+\infty}
          \frac{\partial q_k}{\partial\alpha}\frac{\partial
            q_k}{\partial\rho}\frac{1}{q_k} 
&\sum_{k=0}^{+\infty}
          \left(\frac{\partial q_k}{\partial
              \rho}\right)^2\frac{1}{q_k} 
	 \end{pmatrix}\;.
\end{equation*}
Just like the series $\sum q_k$ itself, the series defining
$I(\alpha,\rho)$ converge very slowly: depending on $\rho$, 
hundreds of thousand terms may be  necessary to get an acceptable
precision, unless a convergence
acceleration method is used. Fortunately the partial sums
increase, so that when computing the inverse $I^{-1}(\alpha, \rho)$ 
the sum of the first $m$ terms yields conservative
confidence intervals; yet we do not consider it satisfactory. 
\vskip 2mm
An obvious answer to the instability problem is to Winsorize the sample 
(see e.g. \cite{Wilcox12}). 
Here this consists of replacing any value of the
sample that pass a certain bound, by the bound itself.
It seems to have been adopted by experimentalists: in the largest
fluctuation experiment ever conducted,  \cite{Boeetal94}
had 4 data above 512, and they did not give a precise count for
them. Indeed, Winsorization outputs acceptable estimates, as long as
the valu²e of $\alpha$ remains small. 
We argue that such a limitation is not acceptable. 
On the one hand, progress in cell counting, particularly using flow
cytometry, will probably lead in the near future to
much higher cell counts. On the other hand, the limitation in
the size of data is a theoretical paradox. For a given type of
cell, increasing the initial number of cells leads to a
proportional increase of $\alpha$. The probabilistic translation is that
LD$(\alpha,\rho)$, as any other compound Poisson distribution, is
infinitely divisible. One could think that increasing $\alpha$
should lead to a more precise estimate of $p$. Due to the heavy tail
property, this is only true if $\rho>1$. In any case, increasing
$\alpha$ is possible only if estimates
can be computed for samples with very high values. As an
example, figure \ref{fig:histogram} shows a histogram of a $10^5$-size
sample of the
LD$(50,0.5)$, on a logarithmic scale. The range of values was
$[128, 1.32 \times 10^{15}]$, the quartiles were $2292$, $6798$, and $30737$. 
There is no hope to calculate the ML estimates on such a sample,
even Winsorized.

\section{Generating Function estimators}
\label{GF}
The idea of using the probability generating function to estimate the
parameter of
a compound Poisson distribution is not new: see 
\cite{RemillardTheodorescu00} and
\cite{Marchesellietal08}. It
turns out that for 
the LD$(\alpha,\rho)$, GF estimators 
are quite comparable in precision to ML
estimators, with a much broader range of calculability, a better
numerical stability, and a negligible
computing time. They are described in this section, and we refer the
reader to the R functions that have been made available on line for
further experiments: 
they include estimation, confidence region, and testing procedures.
\vskip 2mm
Let $(X_1,\ldots,X_n)$ be a sample of independent identically 
distributed random variables, each with probability generating function 
$g_{\alpha,\rho}$. Define the
empirical probability generating function $\hat{g}_n(z)$ as:
$$
\hat{g}_n(z) = \frac{1}{n} \sum_{i=1}^n z^{X_i}\;.
$$
The random variables $z^{X_i}$ are bounded and mutually independent:
by the strong law of large numbers, $\hat{g}_n(z)$ is a
strongly consistent estimator of $g_{\alpha,\rho}(z)$, for any $z$ in
$[0,1]$. Estimates of $\alpha$ and $\rho$ can be obtained by solving
$\hat{g}_n(z)=g_{\alpha,\rho}(z)$ for different values of $z$. 
The implementation is described below.
\vskip 2mm\noindent
Recall the probability generating function of the LD$(\alpha,\rho)$:
$$
g_{\alpha,\rho}(z)=\exp(\alpha (h_\rho(z)-1))\;,
$$ 
with:
$$
h_\rho(z) = \rho z\int_0^1\frac{v^\rho}{1-z+z v}\,\dd v\;.
$$
The derivative of $h_\rho(z)$ with respect to $\rho$ will be
denoted by $h^1_\rho(z)$.
$$
h^1_\rho(z) =
\frac{\partial h_\rho(z)}{\partial\rho}
= z\int_0^1\frac{v^\rho(1+\rho\log(v))}{1-z+z v}\,\dd v\;.
$$
All these functions are easily computed by standard
numerical procedures. For $0<z_1<z_2<1$, consider the following
ratio:
$$
f_{z_1,z_2}(\rho)=\frac{h_\rho(z_1)-1}{h_\rho(z_2)-1}\;.
$$
The function that maps $\rho$ onto 
$y=f_{z_1,z_2}(\rho)$ is continuous and strictly monotone, hence
one-to-one. Therefore the inverse, that maps $y$ onto
$\rho=f^{-1}_{z_1,z_2}(y)$, is well
defined. It is, as well, easily computed by standard numerical methods.
\vskip 2mm
For $0<z_1<z_2<1$, let $\hat{y}_n(z_1,z_2)$ denote the following log-ratio.
$$
\hat{y}_n(z_1,z_2) = 
\frac{\log(\hat{g}_n(z_1))}{\log(\hat{g}_n(z_2))}\;.
$$
An estimator of $\rho$ is obtained by:
$$
\hat{\rho}_n(z_1,z_2) = f^{-1}_{z_1,z_2}(\hat{y}_n)
$$
Then an estimator of $\alpha$ by:
$$
\hat{\alpha}_n(z_1,z_2,z_3) = \frac{\log(\hat{g}_n(z_3))}
{h_{\hat{\rho}_n(z_1,z_2)}(z_3)-1}\;,
$$
where $z_3\in(0\,;1)$ is a new control, possibly different from
$z_1$ and $z_2$. 
They will be referred to as Generating Function (GF) estimators.
By the strong law of large numbers, as $n$ tends to infinity, the random vector 
$(\hat{g}_n(z_1),\hat{g}_n(z_2),\hat{g}_n(z_3))$
converges a.s. to
$(g_{\alpha,\rho}(z_1),g_{\alpha,\rho}(z_2),g_{\alpha,\rho}(z_3))$. Since
the GF estimators  are continuous functions of 
$(\hat{g}_n(z_1),\hat{g}_n(z_2),\hat{g}_n(z_3))$, the following limits hold with
probability $1$.
$$
\lim_{n\to\infty} 
\hat{\rho}_n(z_1,z_2)
=\rho\;,\quad
\lim_{n\to\infty} 
\hat{\alpha}_n(z_1,z_2,z_3)
=\alpha\;.
$$
Therefore the GF estimators are strongly consistent.
Observe that $\hat{\alpha}_n(z_1,z_2,z_3)$ depends on
$\hat{\rho}_n(z_1,z_2)$, whereas
$\hat{\rho}_n(z_1,z_2)$ only depends on the arbitrary choice of the
couple $(z_1,z_2)$. 
\vskip 2mm
Of course the question arises of the variance of the GF
estimators and of their use in hypothesis testing, 
i.e. of computing confidence regions.
Asymptotic variances are obtained through the central limit
theorem, applied to the vector 
$(\hat{g}_n(z_1),\hat{g}_n(z_2),\hat{g}_n(z_3))$. Indeed,
$$
\sqrt{n}\Big(\,
(\hat{g}_n(z_1),\hat{g}_n(z_2),\hat{g}_n(z_3)) - 
(g_{\alpha,\rho}(z_1),g_{\alpha,\rho}(z_2),g_{\alpha,\rho}(z_3))\,\Big)
$$
converges in distribution to the trivariate centered normal
distribution, with covariance matrix
$C=(c(z_i,z_j))_{i,j=1,2,3}$, where:
\begin{equation}
\label{czz}
c(z_i,z_j)=g_{\alpha,\rho}(z_iz_j)-g_{\alpha,\rho}(z_i)g_{\alpha,\rho}(z_j)\;.
\end{equation}
\cite[Proposition 3.1]{RemillardTheodorescu00} give a stronger result,
stating the functional
convergence of $\hat{g}_n(z)$ to a Gaussian process. 

The
following proposition is deduced from Slutsky's theorem, in the
formulation of 
\cite[Theorem 3.4]{RemillardTheodorescu00}. Details of the
calculation are given in the appendix.
\begin{proposition}
\label{prop:clt}
For any $z_1,z_2,z_3$ in $(0\,;1)$ such that 
$z_1\neq z_2$, the couple of random variables 
$$
\sqrt{n}\Big(\,(\hat{\alpha}_n,\hat{\rho}_n)-
(\alpha,\rho)\,\Big)
$$
converges in distribution to the bivariate centered normal distribution with
covariance matrix ${M^t}CM$, where
$M=\left(\begin{array}{cc}A_1&R_1\\A_2&R_2\\A_3&R_3\end{array}\right)$, with
\begin{enumerate}
\item
$
R_1=
\frac{h_\rho(z_2)-1}{\alpha g_{\alpha,\rho}(z_1)
((h_\rho(z_2)-1)h^1_\rho(z_1)-(h_\rho(z_1)-1)h^1_\rho(z_2))}\;,
$
\item
$
R_2=
\frac{h_\rho(z_1)-1}{\alpha g_{\alpha,\rho}(z_2)
((h_\rho(z_1)-1)h^1_\rho(z_2)-(h_\rho(z_2)-1)h^1_\rho(z_1))}\;,
$
\item
$
R_3=0\;,
$
\item
$
A_1=\frac{\alpha h^1_\rho(z_3)}{1-h_\rho(z_3)}R_1\;,
$
\item
$
A_2=\frac{\alpha h^1_\rho(z_3)}{1-h_\rho(z_3)}R_2\;,
$
\item
$
A_3=\frac{1}{g_{\alpha,\rho}(z_3)(h_\rho(z_3)-1)}\;.
$
\end{enumerate}
\end{proposition}
Using proposition \ref{prop:clt} to compute 
confidence regions and intervals, or p-values for hypothesis testing
is standard and we shall not develop that aspect (see e.g.
\cite{Anderson03}). 
\vskip 2mm
The GF estimators such as they have been described so far, 
depend on the three arbitrary values of $z_1$, $z_2$ and $z_3$. Another
tuning parameter will be added. In the LD$(\alpha,\rho)$
the parameter $\rho$ (the heavy tail exponent) 
determines the size and frequency of much larger
values than usual (called ``jackpots'' in \cite{LuriaDelbruck43}). 
For $\rho<1$, some very large values can be
obtained, even for a small $\alpha$. Using the empirical probability
generating function  is a simple way to damp down jackpots, 
and get robust estimates. The
variable $z$ can be seen as a tuning parameter for the damping. 
At the limit case $z=0$, 
$\hat{g}_n(0)$ is simply the frequency of null values, 
and $\hat{\alpha}_n(0)=-\log(\hat{g}_n(0))$ is the
so called $p_0$-estimator of $\alpha$, already considered by 
\cite{LuriaDelbruck43} (it does not depend on
$\rho$). For $z_1=0.1$,
only small observations will be taken into account, whereas for
$z_2=0.9$, much larger values will influence the sum: 
$0.9^{174}\simeq 0.1^{8}$. Thus the empirical probability generating function
damps down jackpots in a differential way according to $z_1$ and
$z_2$. Choosing $z_1=0.1$ and 
$z_2=0.9$ will contrast small values compared to jackpots,
which explains why $\hat{\rho}_n$ can efficiently estimate
$\rho$ for small $\alpha$'s. 
However, for large values of $\alpha$ (say $\alpha>5$), even
$z_2=0.9$ will output very small values, below the machine
precision. This will make the estimates numerically unstable. A
natural way to stabilize them is to rescale the sample, 
dividing all values by a common factor $b$. This amounts to replacing
$z$ by $z^{1/b}$ in the definition of $\hat{g}_n(z)$:
$$
\frac{1}{n}\sum_{i=1}^n z^{X_i/b} =
\frac{1}{n}\sum_{i=1}^n (z^{1/b})^{X_i}=\hat{g}_{n}(z^{1/b})\;. 
$$
We propose to set $b$ to the $q$\textsuperscript{th} quantile of the
sample, where $q$ is another control. In theory,
$z_1,z_2,z_3,q$ should be chosen so as to minimize the asymptotic
variances from proposition \ref{prop:clt}. 
Numerical evidence showed
little dependence of asymptotic variances on the choice
of the tuning parameters. Also, the optimal values
depend on the (unknown) values of $\alpha$ and $\rho$. We
have run simulation experiments from 1000 size samples of size 100 of the
LD$(\alpha,\rho)$, looking for those values of $z_1,z_2,z_3,q$ that
minimized the Mean Squared Error (MSE); 
this was repeated for values
of $\alpha$ from $0.5$ to $5$ and $\rho$ ranging from $0.5$ to
$2$. Our best compromise is $z_1=0.1$, $z_2=0.9$, $z_3=0.8$, $q=0.1$.
In our implementation of the GF estimators, the scaling
factor $b$ is set to the $q$\textsuperscript{th} quantile of the
sample, and all data are divided by that scaling factor (which
amounts to replacing $z_1,z_2,z_3$ by $z_1^{1/b},z_2^{1/b},z_3^{1/b}$). The
estimators $\hat{\alpha}_n$ and $\hat{\rho}_n$ are computed with these
values. We are quite aware that in doing so, 
$z_1^{1/b}, z_2^{1/b}, z_3^{1/b}$ become functions
of the sample, hence the theoretical result of proposition
\ref{prop:clt} does not apply. Nevertheless, the simulation experiments
showed a correct match between theoretical confidence regions and the
experimental ones, so we have used them for computing confidence
intervals and p-values. Observe rescaling of the sample does not
improve the ML method in any way.  
\vskip 2mm 
GF estimates are good, even for
extreme values. As an example, on the sample of size 100.000 of the
LD$(50,0.5)$ from figure \ref{fig:histogram}, the 95\% confidence intervals were
$(48.2\,;\,51.7)$ for $\alpha$ and
$(0.49\,;\,0.51)$ for $\rho$.
On the many repetitions of simulation experiments that we have made,
we have consistently observed the following:
\begin{itemize}
\item GF estimators output, in virtually null computer time,
  reliable values even in cases where the ML estimators fail (large
  $\alpha$, small $\rho$);
\item on samples where both GF and ML estimates can be computed, the
results are quite close;
\item  ML estimates (when they can be computed, i.e. for small
values of $\alpha$), are slightly better 
than the GF ones, so they should definitely be preferred. 
\end{itemize}
However,
initializing the optimization procedure of the ML estimators by a
close enough starting point, is both a guaranty of numerical
stability and economy in computer time. So we have made the obvious
choice of initializing our ML procedures by GF estimates.
\section{Simulation study}
\label{Sim}
Experimental results are reported in this section.
We first checked on published data sets that the ML and GF methods give
coherent results: results are reported in table \ref{tab:published}.
\cite{LuriaDelbruck43} (table~2, p.~504) had data
under 3 different experimental conditions. We have grouped in
sample A experiments numbers 1, 10, 11 and 21b; in sample B
experiments 16 and 17. We have also used data published 
in \cite{Boeetal94,RoscheFoster00,Zheng02}. For each data set
the ML and GF 95\% confidence intervals on
$\alpha$ and $\rho$ are given. The data set from 
\cite{RoscheFoster00} has a high frequency of zeros,
and no jackpot; this explains why $\rho$ cannot be reliably estimated. 
For the data sets from \cite{Boeetal94} and from 
\cite{LuriaDelbruck43} B, the value of
$\rho$ is significantly smaller than $1$: this is a strong argument in
favor of jointly estimating $\alpha$ and $\rho$ rather than assuming
$\rho=1$ as in most fluctuation analyses studies so far.
\begin{table}
\begin{tabular}{|lc|c|c|}
\hline
Reference&size&$\alpha$& $\rho$\\\hline
\cite{LuriaDelbruck43} A&42
& $\begin{array}{c}(5.24\,; 8.74)  \\(5.22\,; 8.89)\end{array}$
& $\begin{array}{c}(0.83\,; 1.33)\\(0.82\,; 1.35)\end{array}$\\\hline
\cite{LuriaDelbruck43} B&32
& $\begin{array}{c}(0.36\,; 1.00)\\(0.35\,; 1.04)\end{array}$
&  $\begin{array}{c}(0.23\,; 0.84)\\(0.18\,; 0.81)\end{array}$\\\hline
\cite{Boeetal94}&1102
& $\begin{array}{c}(0.65\,; 0.77)\\(0.65\,; 0.77)\end{array}$
& $\begin{array}{c}(0.76\,; 0.92)\\(0.73\,; 0.91)\end{array}$\\\hline
\cite{RoscheFoster00}& 52
&  $\begin{array}{c}(1.00\,; 1.80)\\(1.03\,; 1.98)\end{array}$
& $\begin{array}{c}(1.15\,; 6.22)\\(0.00\,; 12.12)\end{array}$\\\hline
\cite{Zheng02}& 30
&   $\begin{array}{c}(6.76\,; 12.94)\\(6.65\,; 12.78)\end{array}$
& $\begin{array}{c}(0.67\,; 1.11)\\(0.66\,; 1.11)\end{array}$\\\hline
\end{tabular}
\caption{Maximum Likelihood (first line) and Generating Function
  (second line) 95\% confidence intervals on $\alpha$ and $\rho$
 for published data sets.}
\label{tab:published}
\end{table}

In order to assess the relative efficiency of the GF 
method, we simulated 1000 samples of 
size $n=100$ for $\alpha$ in $(1,2,4,6,8,10)$ and
$\rho$ in $(2,1.5,1,0.8,0.5)$, and calculated on each sample
the estimates of $\alpha$ and $\rho$ by the GF, ML, and 
Winsorized ML methods (the Winsorization parameter was 500);
the output was the Mean Squared Errors (MSEs) on $\alpha$ 
and $\rho$, over the 1000 samples. 
Apart from the extensive
study of \cite{Boeetal94}, usual fluctuation experiment 
samples have size of order a few tens, which motivated our choice 
for the sample size. The range of values for $\rho$ covers all 
practical situations. For $\alpha$, very small values were not considered
as significant: if $\alpha<1$, a large part of the information is
contained in the frequency of zeros: the $p_0$-method gives 
almost as good results on $\alpha$ as the ML or GF methods. For $\alpha>10$,
the ML estimator fails in most case and its Winsorized version is strongly
biased. The results are reported in table \ref{tab:compareMLa}
(MSEs on estimates of $\alpha$) and table
\ref{tab:compareMLr} (MSEs on estimates of $\rho$).
In all cases where the ML estimator can be computed, the errors are
quite comparable. As $\alpha$ increases, and for $\rho<1$, the ML 
method fails more and more often, and the bias of the Winsorized version 
increases.

\begin{table}
\begin{center}
\begin{tabular}{cccccc}
&$\rho=0.5$&$\rho=0.8$&$\rho=1$&$\rho=1.5$&$\rho=2$\\\hline
\multirow{3}{1cm}{$\alpha=1$}
&0.13&0.13&0.13&0.13&0.12\\
&0.09&0.13&0.13&0.13&0.12\\
&0.14&0.13&0.13&0.13&0.12\\\hline
\multirow{3}{1cm}{$\alpha=2$}
&0.23&0.21&0.22&0.22&0.20\\
&--&0.20&0.20&0.21&0.19\\
&0.26&0.21&0.21&0.21&0.19\\\hline
\multirow{3}{1cm}{$\alpha=4$}
&0.45&0.40&0.39&0.39&0.38\\
&--&0.43&0.39&0.35&0.32\\
&0.63&0.40&0.37&0.34&0.32\\\hline
\multirow{3}{1cm}{$\alpha=6$}
&0.73&0.62&0.56&0.51&0.50\\
&--&0.71&0.58&0.46&0.45\\
&1.38&0.62&0.53&0.46&0.45\\\hline
\multirow{3}{1cm}{$\alpha=8$}
&1.00&0.75&0.70&0.63&0.61\\
&--&1.59&0.80&0.60&0.58\\
&2.55&0.79&0.68&0.60&0.58\\\hline
\multirow{3}{1cm}{$\alpha=10$}
&1.30&0.96&0.88&0.73&0.72\\
&--&--&1.09&0.71&0.68\\
&4.28&1.10&0.87&0.69&0.68\\\hline
\multirow{3}{1cm}{$\alpha=12$}
&1.61&1.15&1.01&0.86&0.82\\
&--&--&1.23&0.84&0.78\\
&6.81&1.37&1.00&0.83&0.78\\\hline
\end{tabular}
\end{center}
\caption{Mean Squared Errors on
  estimates of $\alpha$ (GF on first line, ML
  on second line and Winsorized ML on third line), for
  1000 samples of size 100 of the LD$(\alpha,\rho)$.
  For $\rho=0.5$, ML estimates could not be computed for $\alpha>1$.} 
 \label{tab:compareMLa}
\end{table}

\begin{table}
\begin{center}
\begin{tabular}{cccccc}
&$\rho=0.5$&$\rho=0.8$&$\rho=1$&$\rho=1.5$&$\rho=2$\\\hline
\multirow{3}{1cm}{$\alpha=1$}
&0.09&0.35&0.55&1.09&1.72\\
&0.12&0.39&0.56&1.09&1.73\\
&0.07&0.35&0.55&1.08&1.73\\\hline
\multirow{3}{1cm}{$\alpha=2$}
&0.08&0.33&0.53&1.07&1.64\\
&--&0.38&0.56&1.07&1.61\\
&0.07&0.34&0.53&1.07&1.61\\\hline
\multirow{3}{1cm}{$\alpha=4$}
&0.05&0.32&0.53&1.05&1.62\\
&--&0.39&0.56&1.04&1.59\\
&0.08&0.33&0.53&1.04&1.59\\\hline
\multirow{3}{1cm}{$\alpha=6$}
&0.05&0.32&0.52&1.04&1.58\\
&--&0.41&0.57&1.04&1.57\\
&0.1&0.34&0.53&1.03&1.57\\\hline
\multirow{3}{1cm}{$\alpha=8$}
&0.04&0.31&0.51&1.04&1.57\\
&--&0.43&0.56&1.04&1.56\\
&0.12&0.34&0.52&1.03&1.56\\\hline
\multirow{3}{1cm}{$\alpha=10$}
&0.04&0.31&0.52&1.03&1.56\\
&--&--&0.57&1.04&1.56\\
&0.15&0.34&0.53&1.03&1.56\\\hline
\multirow{3}{1cm}{$\alpha=12$}
&0.02&0.31&0.51&1.02&1.55\\
&--&--&0.58&1.03&1.55\\
&0.18&0.35&0.52&1.02&1.55\\\hline
\end{tabular}
\end{center}
\caption{Mean Squared Errors on
  estimates of $\rho$ (GF on first line, ML
  on second line and Winsorized ML on third line), for
  1000 samples of size 100 of the LD$(\alpha,\rho)$.
  For $\rho=0.5$, ML estimates could not be computed for $\alpha>1$.} 
 \label{tab:compareMLr}
\end{table}
As already said, the GF estimates can be computed for 
much larger values of $\alpha$. Table \ref{tab:largealpha}
shows the MSEs obtained on $\alpha$ and $\rho$ for 
$\alpha$ in $(50,100,150,200)$. As expected, the relative
precision (quotient of the MSE by the value) on $\alpha$  improves as $\rho$
increases, and the relative precision on $\rho$ improves as $\alpha$
increases. Due to the heavy tail property, the relative precision on
$\alpha$ decreases as $\alpha$ increases if $\rho<1$.
We did not try to adapt the GF estimators to much larger values:
we believe that an approximation of the LD$(\alpha,\rho)$ in terms of stable
distributions should be used instead. This is the approach of
\cite{Angerer10} for the case $\rho=1$.
\begin{table}
\begin{center}
\begin{tabular}{cccccc}
&$\rho=2$&$\rho=1.5$&$\rho=1$&$\rho=0.8$&$\rho=0.5$\\\hline
\multirow{2}{1.5cm}{$\alpha=50$}
&3.006    &3.630    &4.764    &6.045   &8.888   \\
&0.161    &0.102    &0.055    &0.045   &0.030   \\ \hline
\multirow{2}{1.5cm}{$\alpha=100$}
&5.966    &8.003    &10.598   &13.036  &22.511  \\
&0.150    &0.100    &0.052   &0.041   &0.030   \\ \hline
\multirow{2}{1.5cm}{$\alpha=150$}
& 9.146   &12.423   &17.395  &20.447   &38.234  \\
& 0.146   & 0.098   &0.051   &0.038    &0.030   \\ \hline
\multirow{2}{1.5cm}{$\alpha=200$}
& 12.992  &18.850   &27.258  &29.854   & 53.048 \\
& 0.150   &0.105    &0.053   &0.038    & 0.030  \\ \hline
\end{tabular}
\end{center}
\caption{Mean Squared Errors on GF
  estimates of $\alpha$ (first line) and $\rho$ (second line), on
  1000 samples of size 100 of the LD$(\alpha,\rho)$.} 
 \label{tab:largealpha}
\end{table}

\begin{figure}
\hspace{-0.2cm}
\begin{tabular}{cccc}
&$\rho=0.5$&$\rho=1$&$\rho=2$\\
$\alpha=1$&
\includegraphics[width=0.2\linewidth]{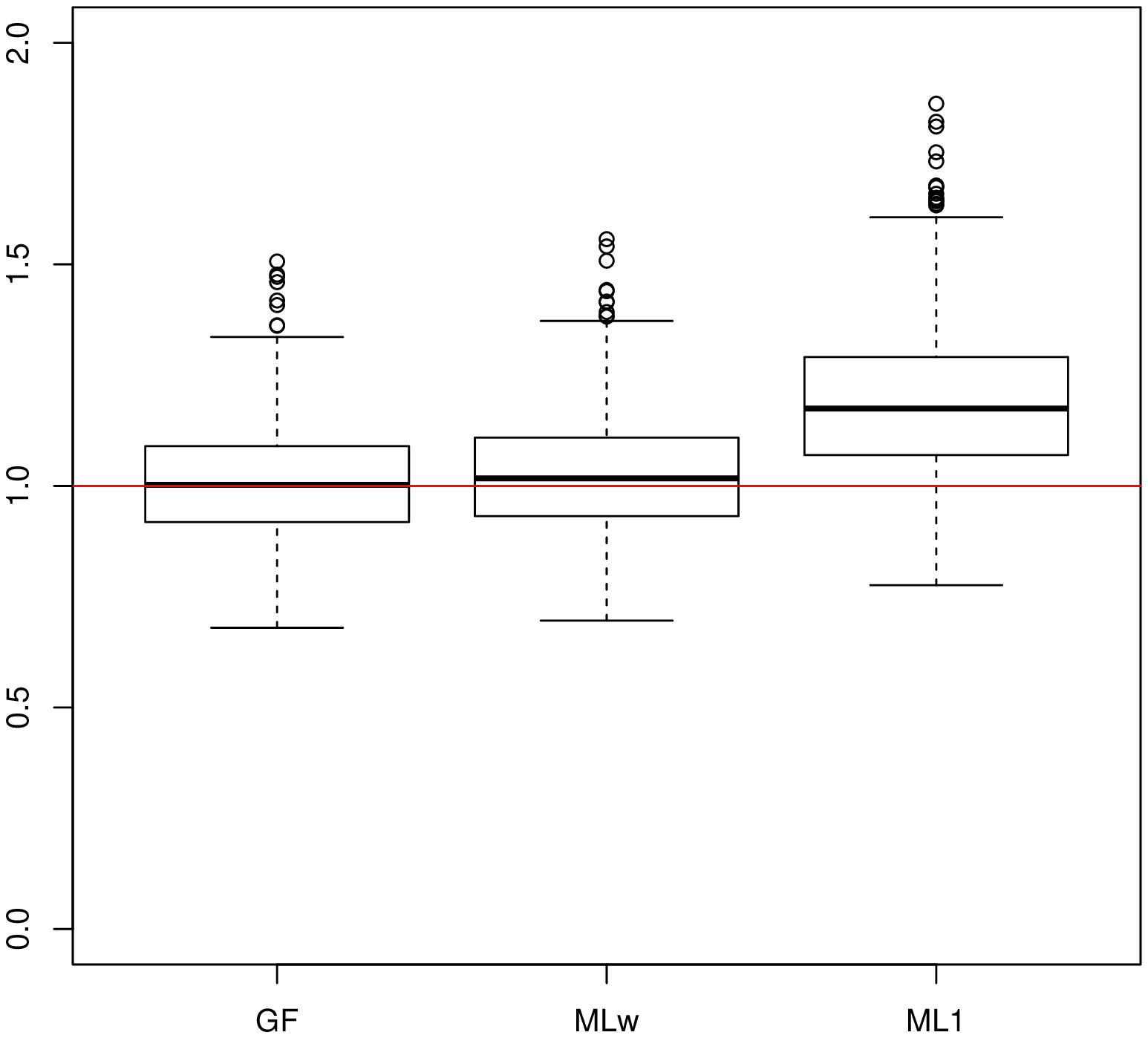}&
\includegraphics[width=0.2\linewidth]{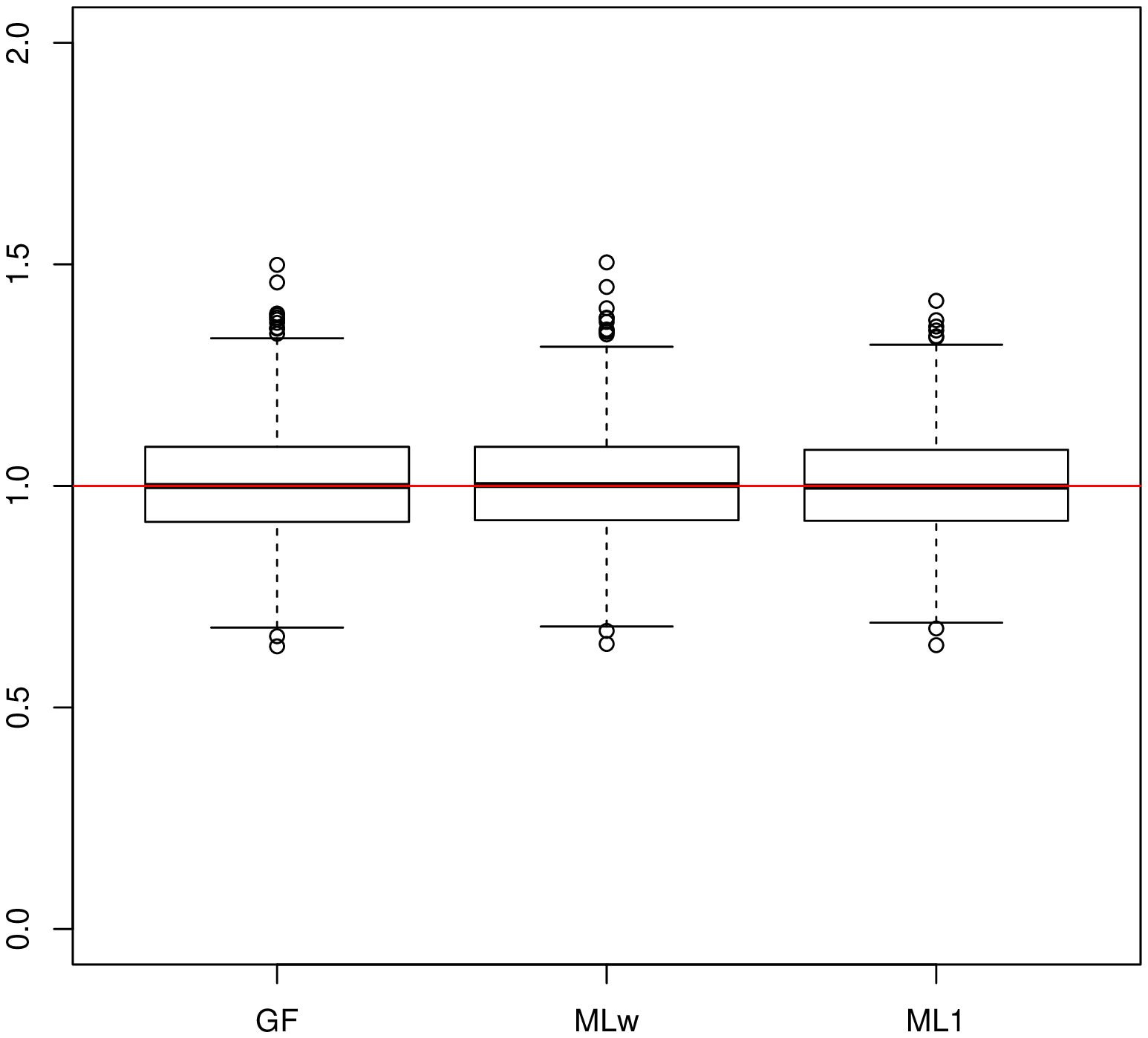}&
\includegraphics[width=0.2\linewidth]{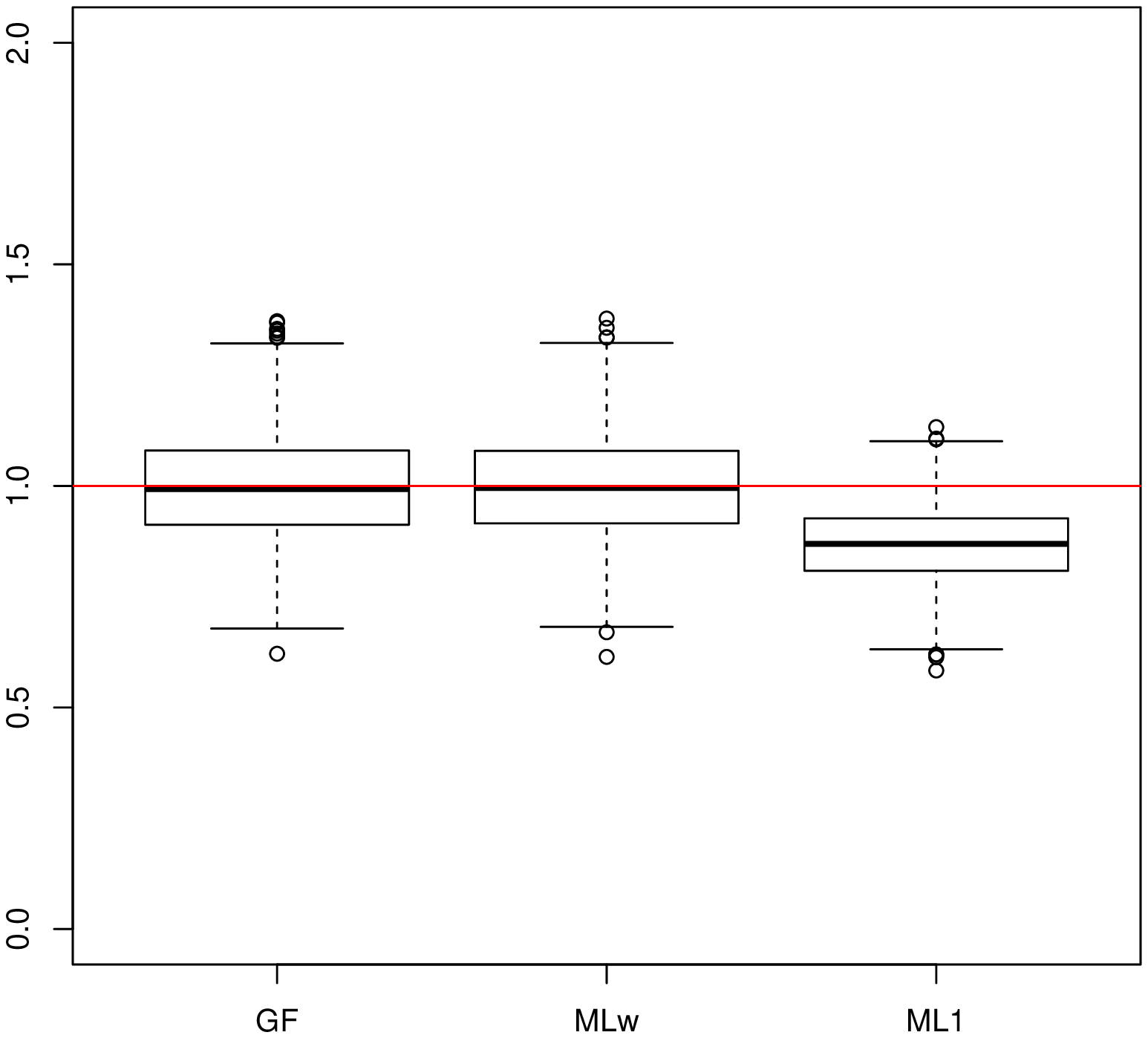}\\
$\alpha=4$&
\includegraphics[width=0.2\linewidth]{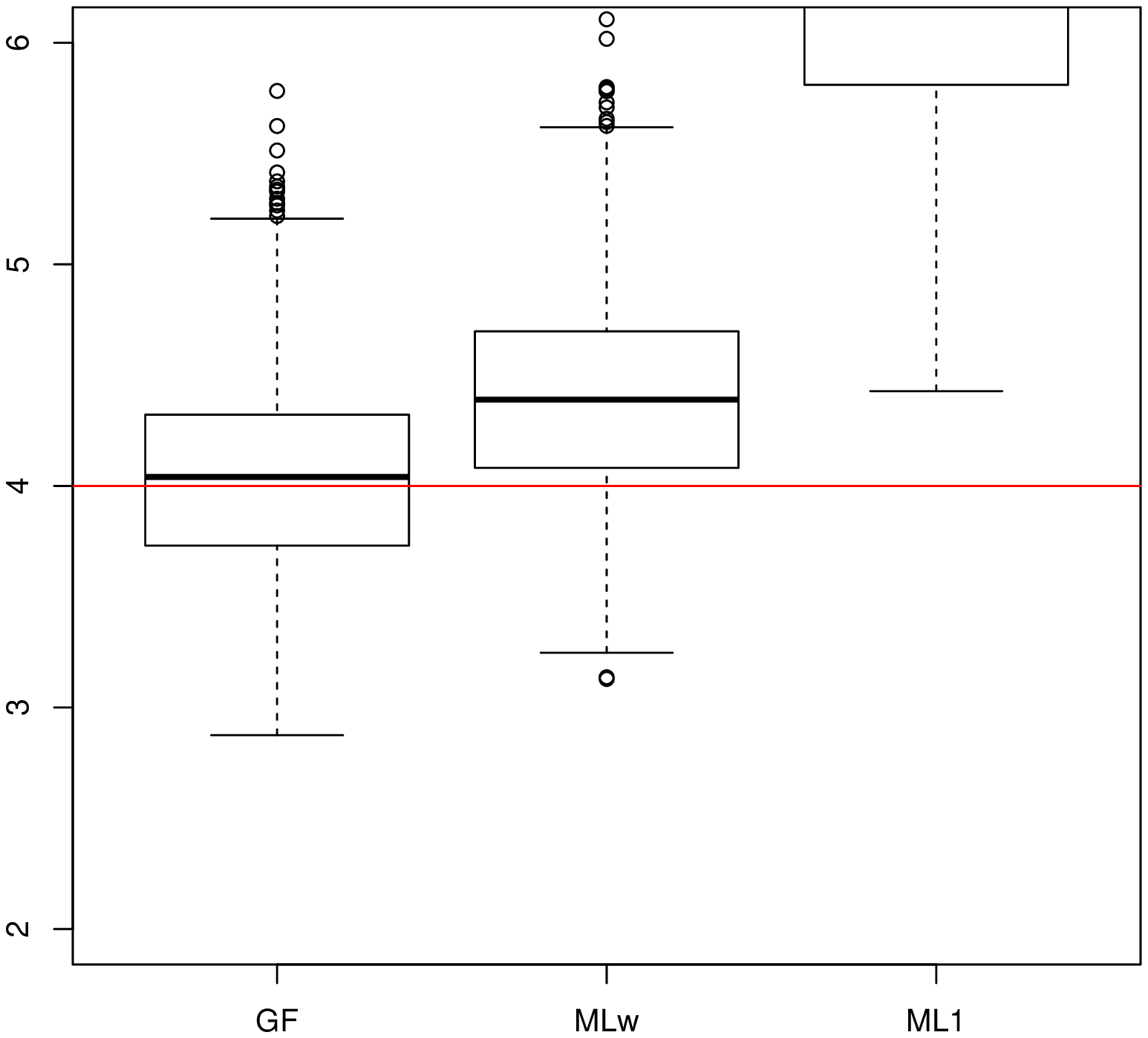}&
\includegraphics[width=0.2\linewidth]{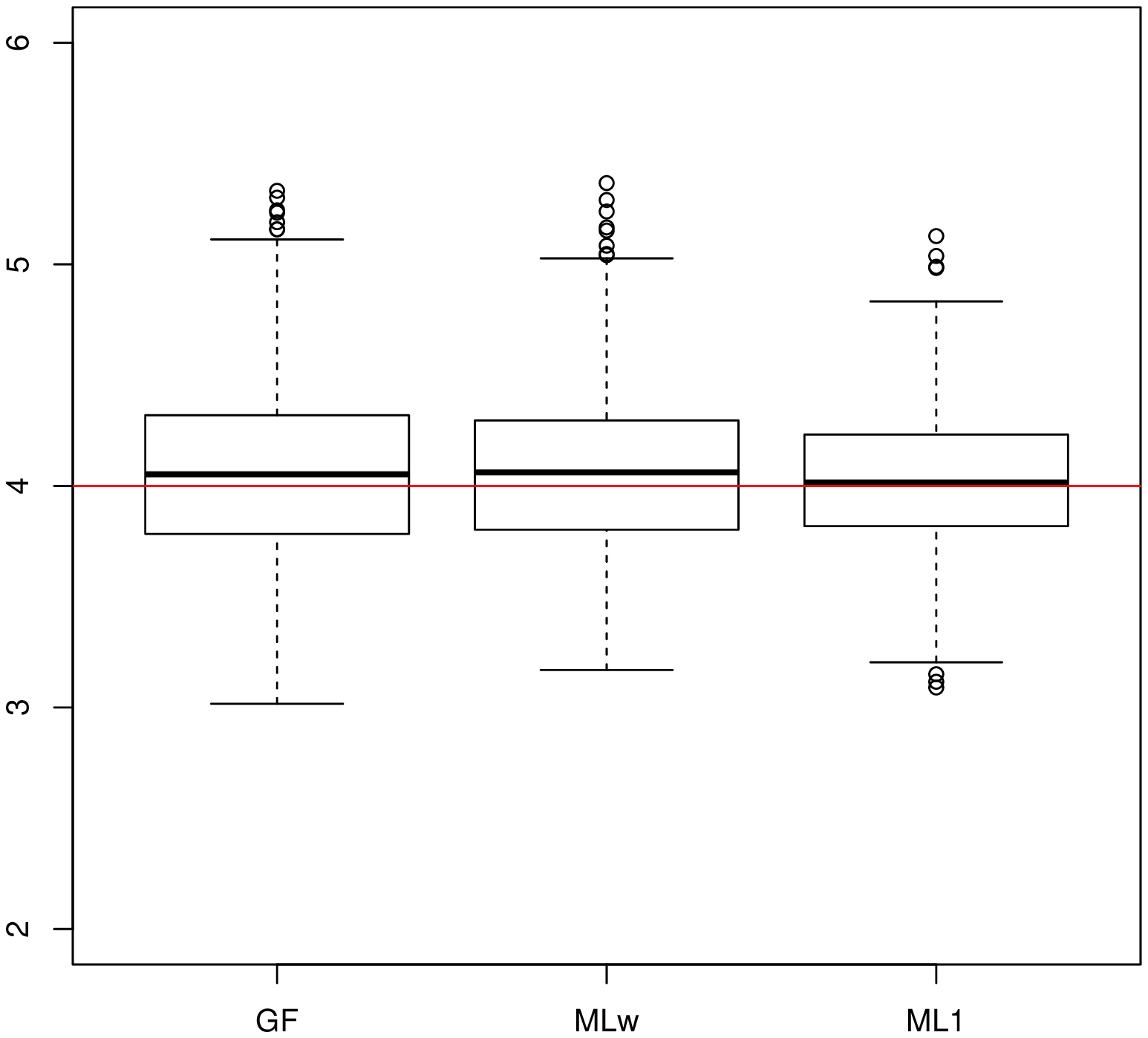}&
\includegraphics[width=0.2\linewidth]{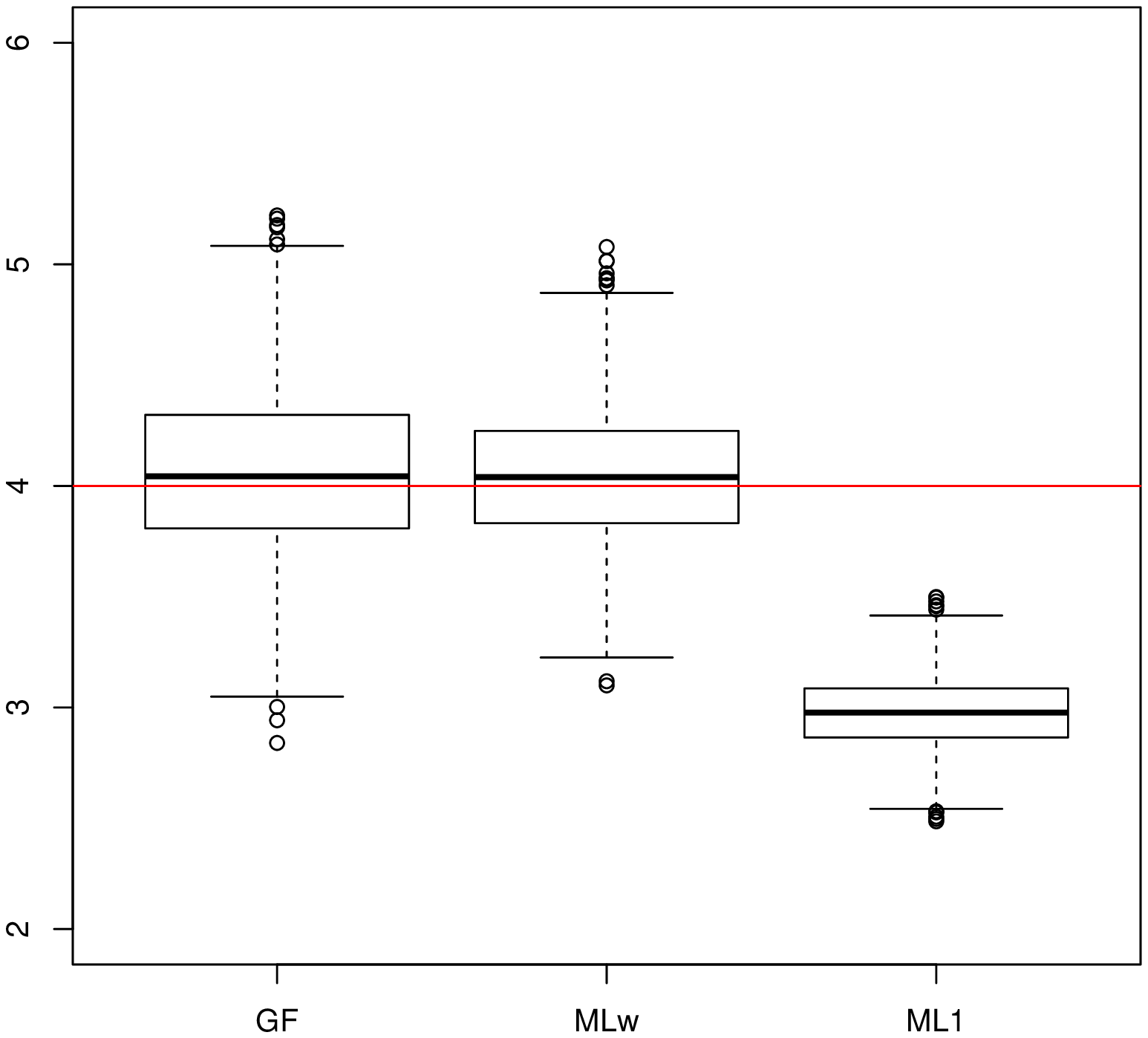}\\
$\alpha=8$&
\includegraphics[width=0.2\linewidth]{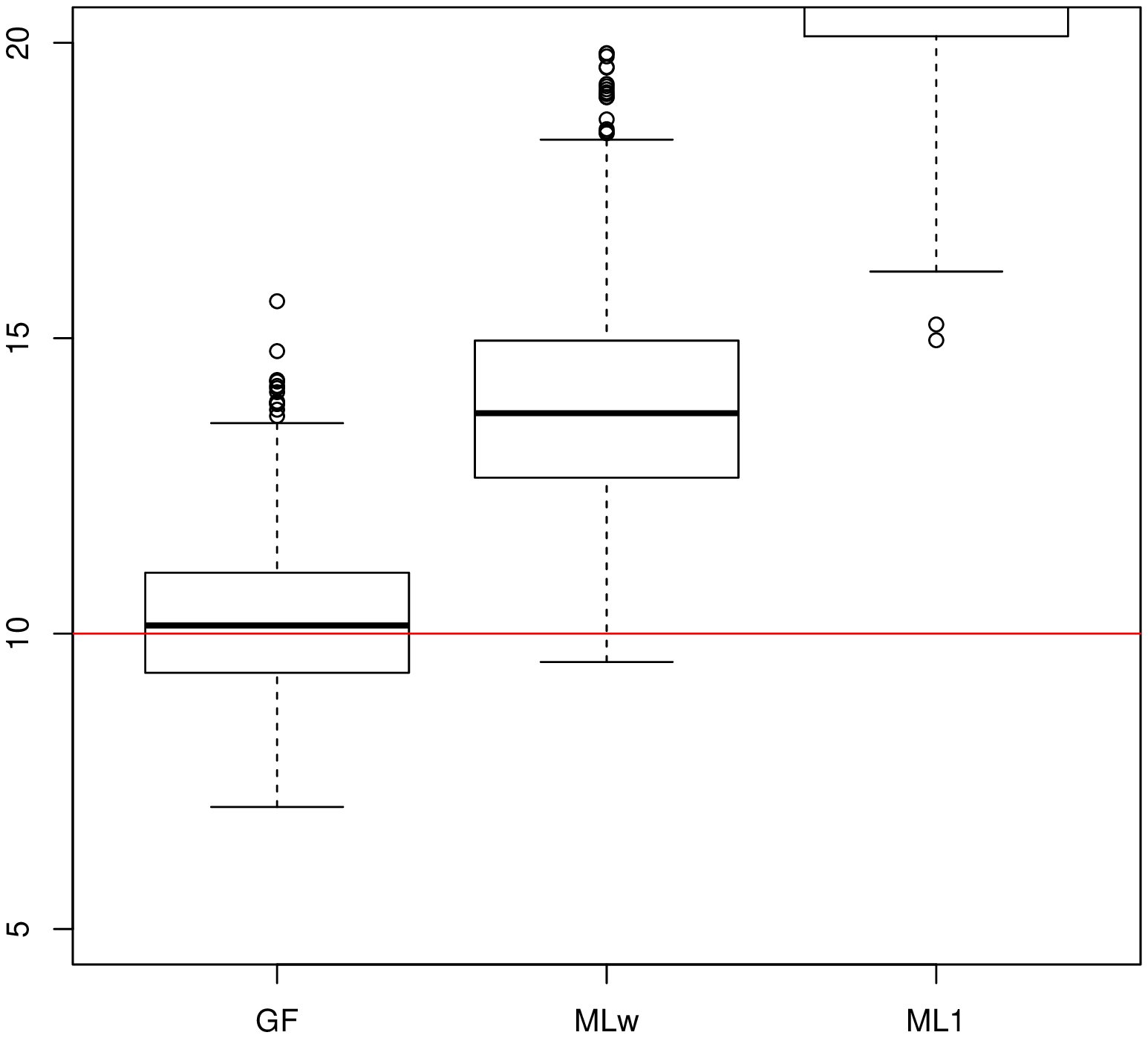}&
\includegraphics[width=0.2\linewidth]{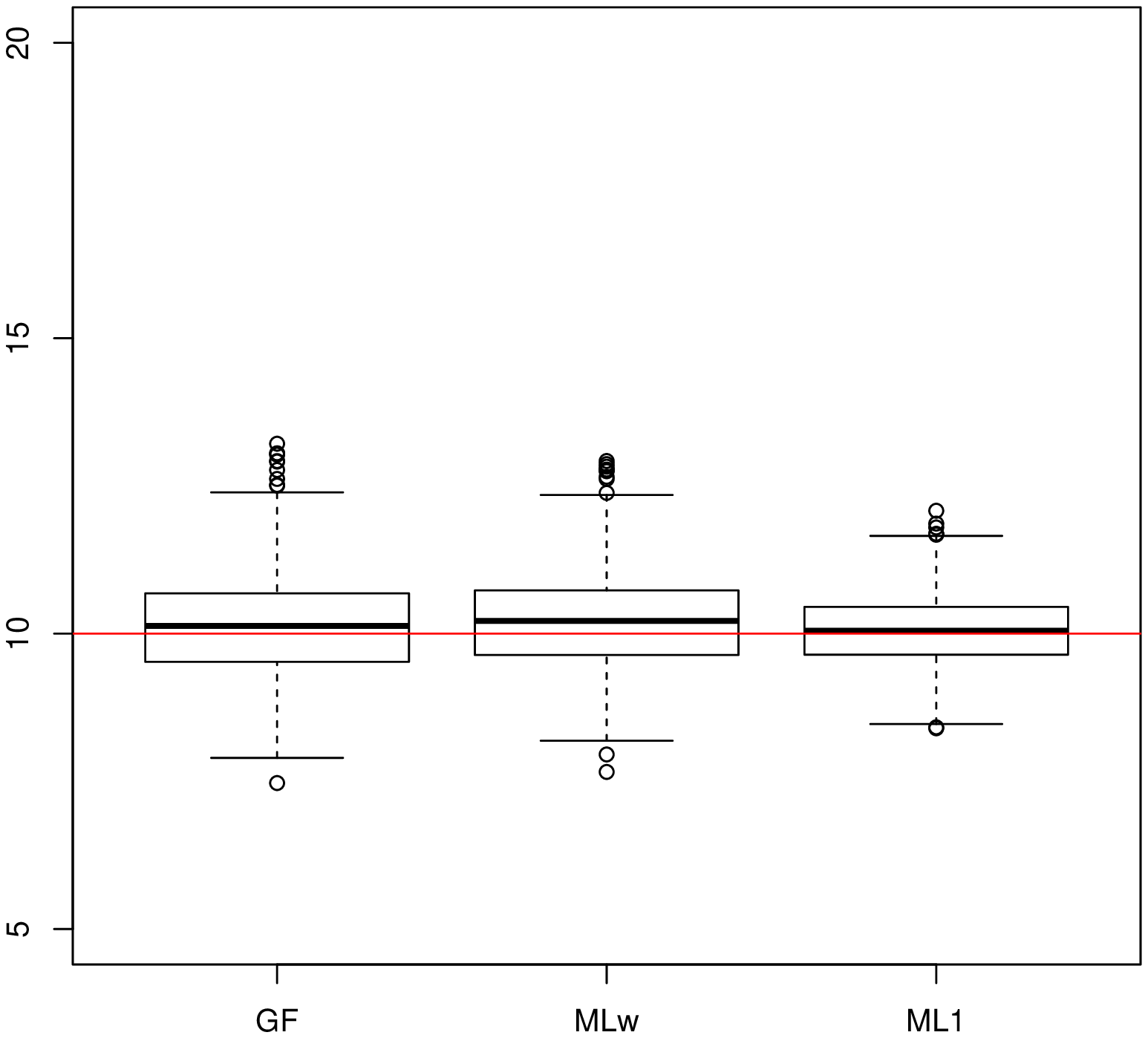}&
\includegraphics[width=0.2\linewidth]{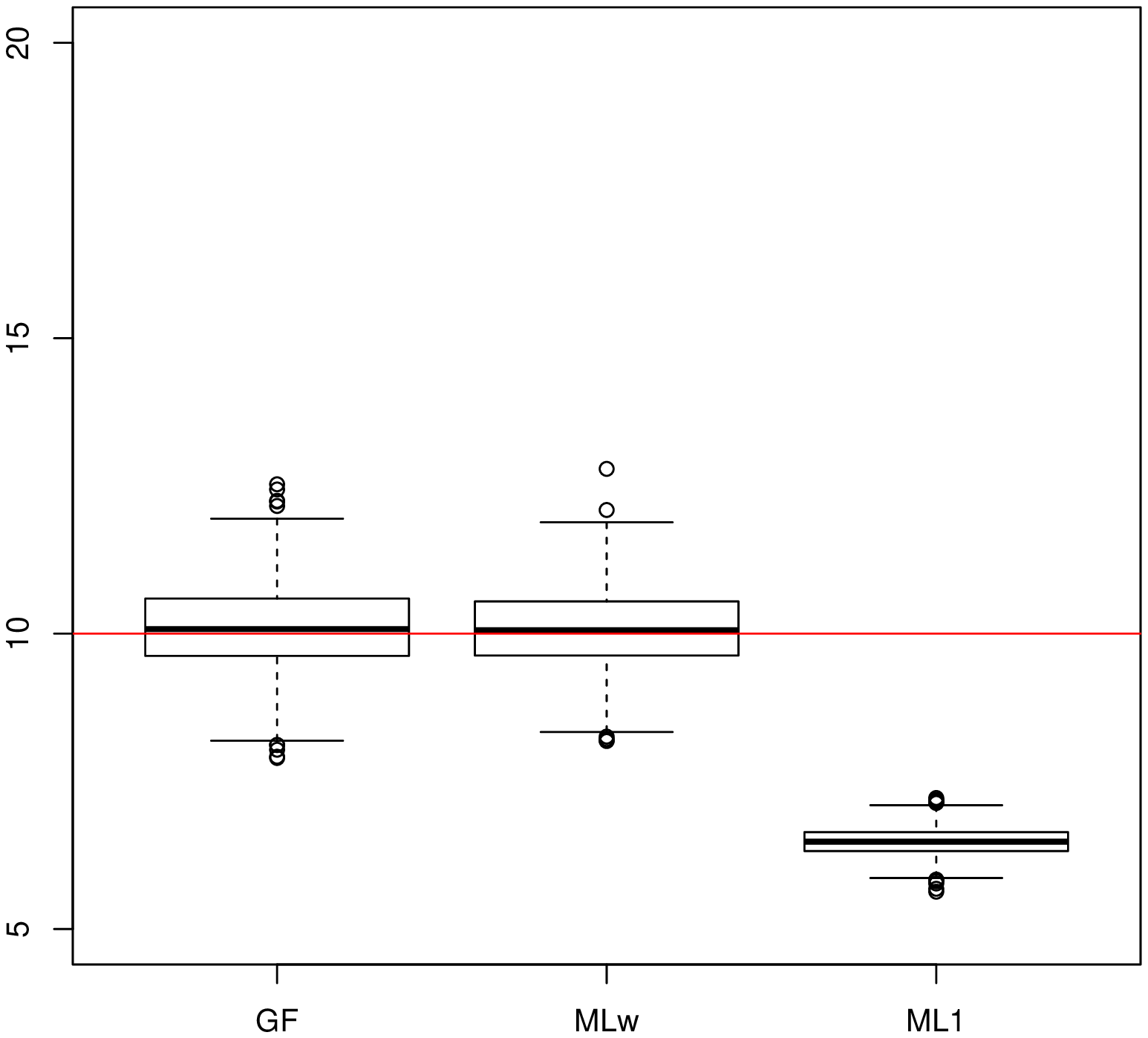}\\
\end{tabular}
\caption{Boxplots of estimates of $\alpha$ based on
  1000 samples of size $100$. The GF estimates (left plot) and 
ML estimates (middle plot) assume the LD$(\alpha,\rho)$ as a model,
the ML$1$ estimate (right plot) assumes the LD$(\alpha,1)$.} 
\label{fig:boxplotsalpha}
\end{figure}

In order to evaluate the bias induced by the equal growth rate
hypothesis, we have compared GF and ML estimates of $\alpha$, to the
ML estimates obtained by assuming the LD$(\alpha,1)$ as a
model. Figure \ref{fig:boxplotsalpha} shows boxplots of
estimates on $1000$ samples of size $100$ of the LD$(\alpha,\rho)$
for $\alpha$ in $(1,4,8)$ and $\rho$ in $(0.5,1,2)$. Not surprisingly
if $\rho=1$, asuming the LD$(\alpha,1)$ as a model slightly improves the
quadratic error. When the true value of $\rho$ is less than $1$, 
assuming $\rho=1$ leads to overestimating $\alpha$ (due to larger and more
frequent jackpots); the larger $\alpha$, the higher the bias. 

\begin{figure}
\centering
\includegraphics[scale=0.65]{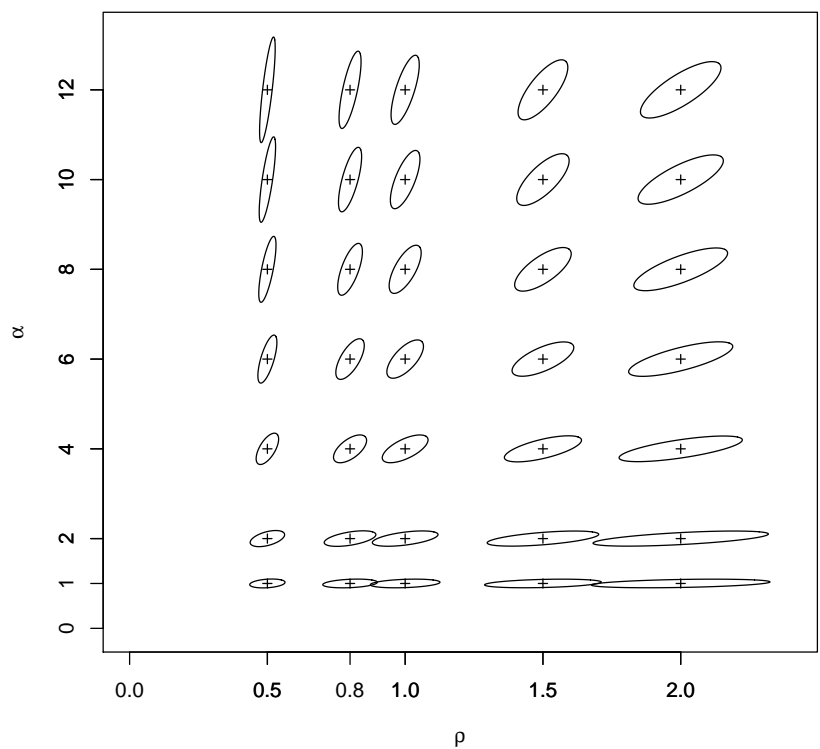}
\caption{Dispersion ellipses for GF estimators of $\alpha$ and $\rho$
  at level 95\%. The
  computation is based on the asymptotic variance 
  of proposition \ref{prop:clt}, for samples of size $100$.}
\label{fig:ellipses} 
\end{figure}

\begin{figure}[!ht]
\centerline{
\includegraphics[width=8cm]{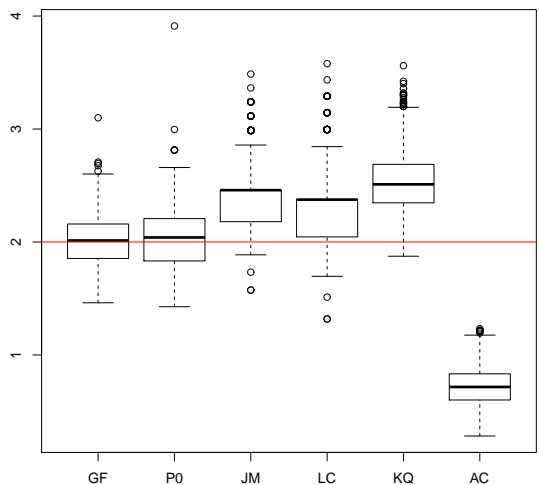}
} 
\caption{Boxplots of estimates of $\alpha$ computed by 6 different
  methods on 1000 samples of size 100 of the LD$(2,0.8)$ distribution.} 
\label{fig:compare}
\end{figure}

The effect of either $\alpha$ or $\rho$ on the estimate of the other
can be seen on the dispersion regions, obtained through
proposition \ref{prop:clt} or through the Fisher information
matrix. Figure \ref{fig:ellipses} displays dispersion ellipses for
different values of $\alpha$ and $\rho$ and samples of size
$100$. When $\alpha$ is small,
the information is concentrated on the value of the distribution at $0$,
which does not depend on $\rho$, hence the bad precision on the estimate
of $\rho$. When $\rho$ is small, frequent jackpots make the
estimates on $\alpha$ less precise. In all cases, the estimates of
$\alpha$ and $\rho$ are positively correlated.
\vskip 2mm
None of the other estimation methods is comparable in quality 
to the ML or GF estimators. We have included in our script a
comparison function for 6 different methods. Figure
\ref{fig:compare} shows a typical output:
boxplots of estimates  of $\alpha=2$, using the
GF estimator and 5 other methods,
computed on 1000 samples of size 100 of the LD$(2,0.8)$
distribution. The 5 other methods are the following (see
\cite{Foster06} for descriptions and references): 
\begin{itemize}
\item[P0:] $p_0$-estimate;
\item[JM:] Lea-Coulson median estimate;
\item[LC:] Jones median estimate;
\item[KQ:] Koch quartiles estimate;
\item[AC:] accumulation of clones estimate.
\end{itemize}
\appendix
\section{Proof of Proposition \ref{prop:clt}}\label{app}
We follow \cite[section 3]{RemillardTheodorescu00}. 
Let $z_1,z_2,z_3$ be three reals in $(0,1)$, two by two distinct. 
Recall that the mapping $f_{z_1,z_2}$, from
$\RR$ to $\RR$, maps $\rho$ onto:
$$
f_{z_1,z_2}(\rho)=\frac{h_\rho(z_1)-1}{h_\rho(z_2)-1}\;.
$$
Define the mapping $\phi$, from $\RR^2$ to $\RR$ by:
$$
\phi(g_1,g_2) = 
f^{-1}_{z_1,z_2}\left(\frac{\log(g_1)}{\log(g_2)}\right)\;,
$$
then $\psi$, from $\RR^3$ to $\RR$ by:
$$
\psi(g_1,g_2,g_3) = \frac{\log(g_3)}{h_{\phi(g_1,g_2)}(z_3)-1}\;.
$$
Finally define $H$, from $\RR^3$ to $\RR^2$, by:
$$
H(g_1,g_2,g_3)=(\psi(g_1,g_2,g_3),\phi(g_1,g_2))\;.
$$
The GF estimator $(\hat{\alpha}_n,\hat{\rho}_n)$ was defined as:
$$
(\hat{\alpha}_n,\hat{\rho}_n) = H(\hat{g}_n(z_1),\hat{g}_n(z_2),\hat{g}_n(z_3))\;.
$$
Applying Theorem 3.4 of \cite{RemillardTheodorescu00}, 
$\sqrt{n}((\hat{\alpha},\hat{\rho})-(\alpha,\rho))$ converges to the
bivariate centered Gaussian vector with covariance matrix $M^tCM$,
where $C$ is the asymptotic covariance of  
$\sqrt{n}((\hat{g}_n(z_1),\hat{g}_n(z_2),\hat{g}_n(z_3))-
(g_{\alpha,\rho}(z_1),g_{\alpha,\rho}(z_2),g_{\alpha,\rho}(z_3)))$,
and $M$ is the Jacobian matrix of $H$, evaluated at 
$(g_{\alpha,\rho}(z_1),g_{\alpha,\rho}(z_2),g_{\alpha,\rho}(z_3))$:
$$
M=
\left(\begin{array}{cc}
\frac{\partial \psi}{\partial g_1}&
\frac{\partial \phi}{\partial  g_1}\\[1ex]
\frac{\partial \psi}{\partial g_2}&
\frac{\partial \phi}{\partial  g_2}\\[1ex]
\frac{\partial \psi}{\partial g_3}&
\frac{\partial \phi}{\partial  g_3}\\[1ex]
\end{array}\right)(g_{\alpha,\rho}(z_1),g_{\alpha,\rho}(z_2),
g_{\alpha,\rho}(z_3))
\;.
$$
The partial derivatives of $\phi$ and $\psi$
in $g_1$, $g_2$, and $g_3$ are computed as follows.
\begin{eqnarray*}
\displaystyle{\frac{ \partial \phi}{\partial g_1}} &=&
\displaystyle{\frac{\partial f^{-1}_{z_1,z_2}(\log(g_1)/\log(g_2))}
{\partial g_1}}\\[2ex]
&=&\displaystyle{
\frac{1}{g_1\log(g_2)}\,\frac{(h_\rho(z_2)-1)^2}
{h^1_\rho(z_1)(h_\rho(z_2)-1)-h^1_\rho(z_2)(h_\rho(z_1)-1)}
}
\end{eqnarray*}
Taking the value at $g_1=g_{\alpha,\rho}(z_1)$ and
$g_2=g_{\alpha,\rho}(z_2)$ gives:
\begin{eqnarray*}
&&\displaystyle{\frac{ \partial \phi}{\partial g_1}
(g_{\alpha,\rho}(z_1),g_{\alpha,\rho}(z_2))}\\[2ex]
&=&\displaystyle{
\frac{1}{g_{\alpha,\rho}(z_1)\alpha(h_\rho(z_2)-1)}\,\frac{(h_\rho(z_2)-1)^2}
{h^1_\rho(z_1)(h_\rho(z_2)-1)-h^1_\rho(z_2)(h_\rho(z_1)-1)}
}\\[2ex]
&=&\displaystyle{
\frac{h_\rho(z_2)-1}{\alpha g_{\alpha,\rho}(z_1)
((h_\rho(z_2)-1)h^1_\rho(z_1)-(h_\rho(z_1)-1)h^1_\rho(z_2))}=R_1\;.}
\end{eqnarray*}
The partial derivative of $\phi$ in $g_2$ is obtained by swapping
indices $1$ and $2$:
\begin{eqnarray*}
&&\displaystyle{\frac{ \partial \phi}{\partial g_2}
(g_{\alpha,\rho}(z_1),g_{\alpha,\rho}(z_2))}\\[2ex] 
&=&\displaystyle{
\frac{h_\rho(z_1)-1}{\alpha g_{\alpha,\rho}(z_2)
((h_\rho(z_1)-1)h^1_\rho(z_2)-(h_\rho(z_2)-1)h^1_\rho(z_1))}
=R_2\;.}
\end{eqnarray*}
The derivative of $\phi$ in $g_3$ is null. The derivative of $\psi$ in
$g_1$ is:
$$
\displaystyle{\frac{\partial \psi}{\partial g_1}}=
\displaystyle{h^1_\rho(z_3)\frac{\log(g_3)}{(h_{\phi(g_1,g_2)}-1)^2}
\frac{\partial \phi}{\partial g_1}\;.}
$$
The value at
$(g_{\alpha,\rho}(z_1),g_{\alpha,\rho}(z_2),g_{\alpha,\rho}(z_3))$
is:
$$
\displaystyle{\frac{ \partial \psi}{\partial g_1}
(g_{\alpha,\rho}(z_1),g_{\alpha,\rho}(z_2),g_{\alpha,\rho}(z_2))} 
=\displaystyle{
\frac{\alpha h^1_\rho(z_3)}{1-h_\rho(z_3)}R_1=A_1\;.}
$$
Replace $ R_1$ by $R_2$ to get $A_2$:
$$
\displaystyle{\frac{ \partial \psi}{\partial g_2}
(g_{\alpha,\rho}(z_1),g_{\alpha,\rho}(z_2),g_{\alpha,\rho}(z_2))} 
=\displaystyle{
\frac{\alpha h^1_\rho(z_3)}{1-h_\rho(z_3)}R_2=A_2\;.}
$$
Finally, the derivative of $\psi$ in $g_3$ is:
$$
\displaystyle{\frac{\partial \psi}{\partial g_3}}=
\displaystyle{\frac{1}{g_3(h_{\phi(g_1,g_2)}(z_3)-1)}\;,}
$$
Hence:
$$
\displaystyle{\frac{ \partial \psi}{\partial g_3}
(g_{\alpha,\rho}(z_1),g_{\alpha,\rho}(z_2),g_{\alpha,\rho}(z_3))} 
=\displaystyle{
\displaystyle{\frac{1}{g_{\alpha,\rho}(z_3)(h_{\rho}(z_3)-1)}=A_3}\;.}
$$

\section*{Acknowledgements}
The authors are grateful to Anestis Antoniadis, Jo\"el Gaff\'e,
Alain le Breton, and Dominique Schneider for helpful 
and pleasant discussions. They are indebted to the two anonymous
referees for insightful suggestions.

\begin{supplement}
\sname{Additional material}\label{suppA}
\stitle{R script}
\slink[url]{http://ljk.imag.fr/membres/Bernard.Ycart/LD/}
\sdescription{Set of R functions for statistical computation with
  Luria-Delbr\"uck distributions, including random sample simulation,
  estimation of parameters with ML and GF methods, asymptotic variance
matrices, confidence intervals and p-values for hypothesis testing.}
\end{supplement}

%

\end{document}